\documentclass[aps,prl,showpacs,letterpaper,twocolumn, floatfix]{revtex4}%
\usepackage{graphicx}
\usepackage{bm}
\usepackage{epsf}
\usepackage{rotating}
\usepackage{epsfig,graphics,rotate,color}
\usepackage{wrapfig}
\usepackage{amssymb}
\usepackage{amsmath}
\usepackage{amsfonts}
\usepackage{array,hhline,dcolumn}%
\usepackage{gensymb}
\usepackage{placeins}
\usepackage[colorlinks=true,citecolor=blue,linkcolor=blue]{hyperref}

\providecommand{\U}[1]{\protect\rule{.1in}{.1in}}
\newcommand\scalemath[2]{\scalebox{#1}{\mbox{\ensuremath{\displaystyle #2}}}}

\begin{document}
\title{First Constraints on the Complete Neutrino Mixing Matrix with a Sterile Neutrino}
\author{G.H. Collin$^1$, C.A. Arg\"uelles$^{1}$, J.M. Conrad$^1$, M.H. Shaevitz$^2$}
\affiliation{$^{1}$ Massachusetts Institute of Technology, Cambridge, MA 02139, USA}
\affiliation{$^{2}$ Columbia University, New York, NY 10027, USA}

\begin{abstract}
Neutrino oscillation models involving one extra mass eigenstate beyond the standard three ($3+1$) are fit to global short baseline experimental data and the recent IceCube $\nu_\mu+\bar\nu_\mu$ disappearance search result.  We find a best fit of $\Delta m^2_{41} = 1.75\; \text{eV}^2$ with $\Delta \chi^2_{null-min}$ (dof) of 50.61 (4).  We find that the combined IceCube and short baseline data constrain $\theta_{34}$ to 
$< 80^\circ (< 6^\circ) $ at 90\% C.L. for $\Delta m^2_{41} \approx 2 (6)~\text{eV}^2$, which is improved over present limits.
Incorporating the IceCube information  provides the first constraints on all entries of the 3+1 mixing matrix.
\end{abstract}


\maketitle

\section{Introduction}

Neutrino oscillations are due to quantum-mechanical effects that occur if the neutrino mass eigenstates are mixtures of the neutrino flavor eigenstates.    These effects have been observed in multiple experiments \cite{pdg}. Most neutrino oscillation data sets fit well into a model involving three active neutrinos that map to three distinct mass states through a unitary mixing matrix \cite{pdg}.   This three-neutrino model, then, has two independent squared-mass splittings, $\Delta m^2_{ji} = m_j^2-m_i^2$. The frequency of vacuum neutrino oscillations depends upon the magnitude of these squared-mass splittings. The larger of the two well-confirmed splittings, historically called the atmospheric splitting, is $\Delta m^2_{atm}=2.3 \times 10^{-3} {\rm eV}^2$ \cite{nufit}, while the smaller well-confirmed splitting, called the solar splitting, is $\Delta m^2_{sol}=7.5 \times 10^{-5} {\rm eV}^2$ \cite{nufit}.  

On the other hand, a set of anomalous experiments provide indications of oscillations with substantially different frequency than the solar and atmospheric results.   Results from  LSND \cite{LSND},  MiniBooNE \cite{nubarminiosc2, miniboonelowe2}, reactor experiments \cite{Bugey, mention}, and Gallium source experiments \cite{SAGE3, GALLEX3} are consistent with a third mass splitting, $\Delta m^2\sim 1$ eV$^2$.  These experiments are generally classified as ``short baseline'' (SBL), meaning that they are designed with a distance-to-energy ratio for the neutrino source and detector of around $L/E\sim 1 {\rm m/MeV}$.  
LSND and MiniBooNE observed electron flavor neutrino appearance in a muon-flavored beam.   The reactor and source experiments observed electron-flavor disappearance.  The significance of these signals ranges from 2$\sigma$ to 4$\sigma$, and hence are less compelling individually than the solar and atmospheric results.  However, taken together, the results point to a new oscillation parameter region.  To accommodate a third squared-mass splitting that is not the sum of the atmospheric and solar splittings, one must introduce a fourth neutrino mass state into the model.
Since LEP Z$^0$ decay measurements are consistent with only three low mass, active neutrinos \cite{zwidth}, an additional fourth neutrino flavor must be inactive and is historically called sterile.     

However, other SBL experiments sensitive to this higher oscillation frequency have observed null results \cite{Karmen, ConradShaevitz, MBNuMI, Kendall, K1, NOMAD1, CCFR84, CDHS, MINOSCC1, MINOSCC3}.  In particular, muon flavor disappearance has not been observed in SBL experiments.   These limits must also be accounted for in any model with extra neutrino flavors.  As a result, global fits of data employing three active neutrinos and one sterile neutrino, called 3+1 models, have a limited allowed range in vacuum oscillation parameter space \cite{Collin:2016rao, kopp_sterile_2013, giunti_pragmatic_2013}.   Nevertheless, 3+1 models that fit all of the data sets do remain and have prompted a suite of new SBL experiments, which are now underway \cite{uboone} or are in design \cite{SBNproposal, IsoDAR, Prospect, sox}.

In Ref.~\cite{Collin:2016rao}, we have reported the results of global fits to the SBL data that yield allowed regions at 90\% CL at three $\Delta m^2$ values at approximately 1, 1.75 and 6 eV$^2$. 
In this paper, we expand these 3+1 fits to include a new, highly restrictive oscillation limit from the IceCube Experiment \cite{IceCubePRL}.   Because the IceCube analysis relies on matter effects rather than vacuum oscillations, this new data set breaks degeneracies, allowing, for the first time, to fill in all of the elements of the 3+1 mixing matrix.

\section{Constraints from SBL Experiments}

SBL experiments have direct sensitivity to neutrino oscillations involving electron and muon flavor neutrinos, but do not have direct sensitivity to transitions involving the tau neutrino flavor.   This is because the $\nu_\tau$ threshold for charged current interactions of 3.4 GeV suppresses charged current interactions for these low energy SBL experiments.
A full 3+1 model, however, has a  $4\times4$ matrix that connects all three active plus single sterile flavor states to the four mass states:
\begin{equation}
U_{3+1} = \begin{bmatrix}
U_{e1} & U_{e2} & U_{e3} & U_{e4} \\ 
U_{\mu 1} & U_{ \mu 2} & U_{\mu 3} & U_{\mu4} \\
U_{\tau 1} & U_{\tau 2} & U_{\tau 3} & U_{\tau4} \\
U_{s1} & U_{s2} & U_{s3} & U_{s4}
\end{bmatrix}. \label{4mixmx}
\end{equation}
Due the high $\tau$ production threshold, the SBL experiments can only directly constrain the elements $U_{e4}$ and $U_{\mu 4}$.

The observed anomalous mass splitting associated with oscillations to the fourth neutrino flavor is very large compared to the solar and atmospheric cases. Thus, for these oscillations, one can make the approximations that $\Delta m^2_{41}\approx \Delta m^2_{42} \approx \Delta m^2_{43}$ and $\Delta m^2_{21} \approx \Delta m^2_{32}\approx 0$. This leads to the SBL approximation for the 
vacuum oscillation probability formula for $\nu_{\alpha} \rightarrow \nu_{\beta}$: 
\begin{equation}
P_{\alpha \beta} = \delta_{\alpha\beta} - 4 (\delta_{\alpha\beta} - U_{\alpha 4} U_{\beta 4}^{*}) U_{\alpha 4}^{*} U_{\beta 4} \sin^{2}\left(\frac{\Delta m_{41}^2 L}{4E} \right).
\label{PintermsofU}
\end{equation}
In this equation, $L$ is the distance the neutrino travels and $E$ is the energy of the neutrino.
For a given choice of flavors $\alpha$ and $\beta$,  this is equivalent to a two neutrino model with a mixing amplitude of 
\begin{equation}
\sin^2{2\theta_{\alpha\beta}} = |4(\delta_{\alpha\beta} - U_{\alpha 4} U_{\beta 4}^{*}) U_{\alpha 4}^{*} U_{\beta 4}|. \label{sin22theq}  
\end{equation}
Thus, in this notation, muon-to-electron flavor appearance experiments measure $\sin^2{2\theta_{\mu e}}$, and the disappearance experiments measure $\sin^2{2\theta_{e e}}$ and $\sin^2{2\theta_{\mu \mu}}$.

This paper makes use of the global fits to SBL data reported in 
Ref.~\cite{Collin:2016rao}.  
The experiments are chosen to be relevant in the range of $\Delta m^2>0.3$ eV$^2$ at 90\% CL, which is the limit of LSND \cite{LSND}.    We fit in the range of $0.1<\Delta m^2<100$ eV$^2$. 
The specifics of the SBL data sets are given in Table 1 of Ref.~\cite{Collin:2016rao}, and the associated text, and so we very briefly explain the choices here.     With respect to electron neutrino appearance, we include LSND \cite{LSND}, MiniBooNE ($\nu$ and $\bar \nu$ from the BNB flux) \cite{miniboonelowe2, miniboonelowe1, nubarminiosc1, nubarminiosc2}, MiniBooNE (NuMI off-axis $\nu$ flux) \cite{NuMIMB}, KARMEN \cite{Karmen} and NOMAD\cite{NOMAD1}.   With respect to electron neutrino disappearance, we include Bugey \cite{Bugey}, the Gallium Experiments \cite{SAGE3, GALLEX3} and the Karmen/LSND cross section analysis \cite{ConradShaevitz}.  With respect to $\nu_\mu$ disappearance, we include the MiniBooNE-SciBooNE joint analyses in $\nu$ and $\bar \nu$ running \cite{Mahn:2011ea, Cheng:2012yy}, the CDHS result \cite{CDHS},  MINOS results from 2006 and 2008 \cite{MINOSCC1, MINOSCC2} that are strictly from charged current analysis, and CCFR84 \cite{CCFR84}.

\begin{figure}[t!]
\center
\includegraphics[width=\columnwidth]{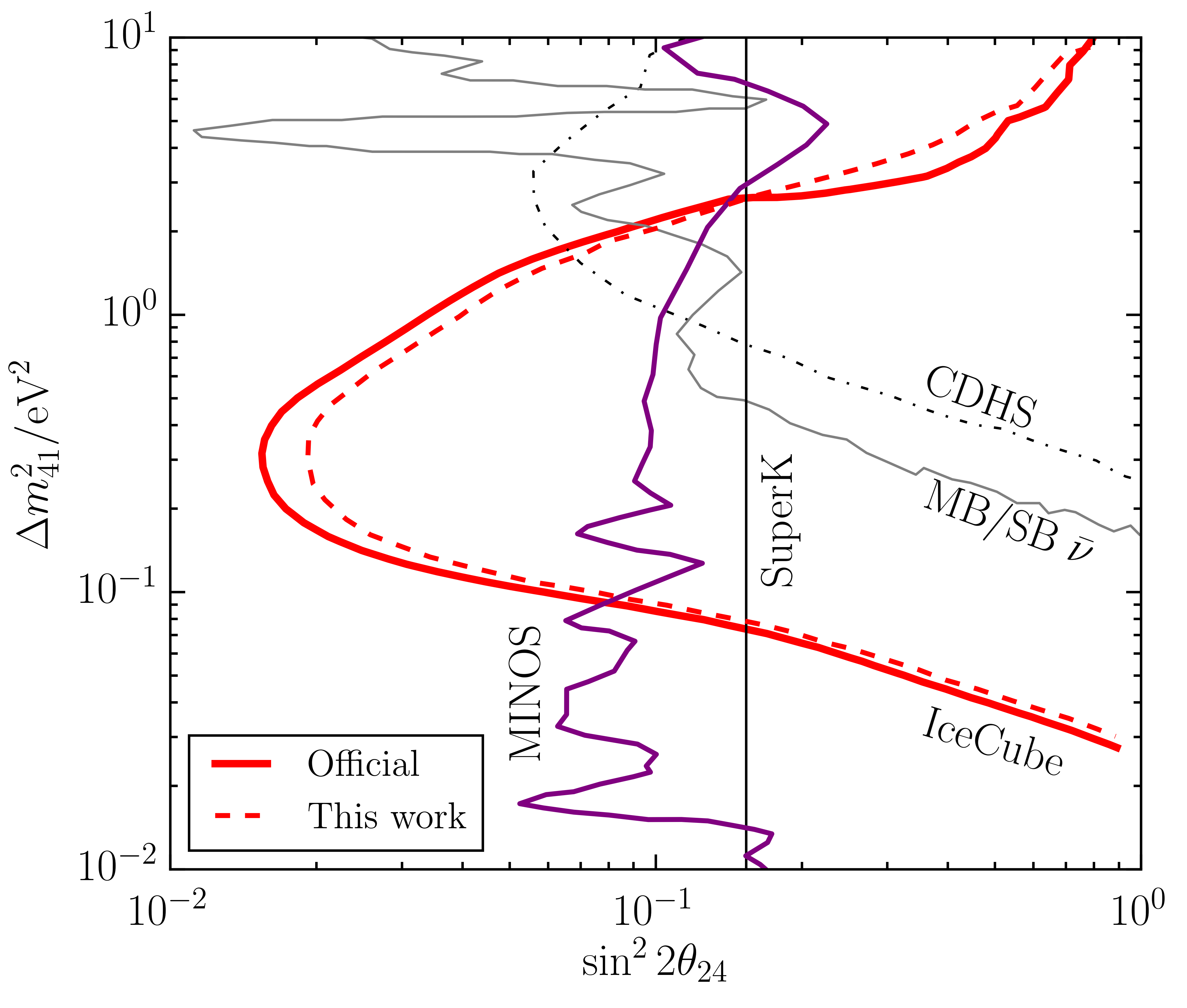} \\
\caption{Comparison of 90\% CL limits for muon flavor disappearance of IceCube 2016, MINOS 2016, CDHS and MiniBooNE-SciBooNE.  Our reconstruction of the IceCube result using the data release is indicated by the dashed line. \\ \label{MINOSIce}}
\end{figure}

There are two results published in 2016 that are not included in these fits.   
The 2016 Daya Bay $\bar\nu_e$ disappearance result, which is published for  $2\times 10^{-4}<\Delta m^2<0.3$ eV$^2$ \cite{An:2016luf}, need not be included in these fits.  The small overlap is in a region dominated by the Bugey result already used in the fit \cite{Bugey}.     The 2016 MINOS $\nu_\mu$ disappearance result \cite{MINOS2016} is not included in the fit for two reasons.   First, this result is not competitive with the IceCube result and other data sets already used in the region of interest for the fits, as can be seen in Fig.~\ref{MINOSIce}.   Second, this disappearance result incorporates neutral current data, with a background subtraction for the relatively large \cite{Minosth13} $\nu_e$ intrinsic flux.  Thus the MINOS limit is dependent on their assumption that $|U_{e4}|^2 = 0$ and cannot be directly used in our global fits that need to include $\nu_e$ disappearance in an unrestricted way. 

These fits also do not include data from cosmology because the CMB and large scale structure (LSS) constraints on the presence of a fourth neutrino are model dependent.  The dependencies include assuming
a ``standard" thermal history for the Universe \cite{KevAnnRev}. Sterile neutrino thermalization can be suppressed a number of plausible ways \cite{Kev2,Dasgupta:2013zpn,Hannestad:2013ana,Bento:2001xi,Chu:2006ua,Foot:1995bm,Gelmini:2004ah,Ho:2012br,Hamann:2011ge}.   In fact,  thermalization may not occur when one considers models with full four-neutrino mixing \cite{Kev1}.    Introducing the assumption that sterile neutrinos have very weak pseudoscalar interactions that are unobservable in the short baseline data not only resolves the apparent disagreement between the 3+1 models and CMB, it also predicts a Hubble constant in agreement with local measurements \cite{Sten}.
Changes in the assumption of the influence of dark energy on the expansion history and growth structure also influences the comological results \cite{KevAnnRev}.   Based on this, rather than include cosmology in the fit, what is most interesting is to fit the cosmological data separately from the oscillation experiments, and then consider the meaning of discrepancies.

The global fit favors a model with one mass state dominated by the sterile flavor.  The three assumed degenerate mass states are dominated by the active flavors, as is demanded by the solar and atmospheric neutrino results.    The SBL fits cannot distinguish the mass hierarchy, that is, whether the dominantly sterile flavor is the highest mass state, which is called a 3+1 hierarchy, or the lowest mass state, which is called a 1+3 hierarchy.

\section{Incorporating IceCube Data}

In this paper, we expand the 3+1 fits to include data from IceCube \cite{IceCubePRL}, which is quite different in design from the SBL experiments.   
It makes use of measurements of
the atmospheric $\nu_\mu$ flux, studied as a function of the zenith angle and energy in the range from 400 GeV to 20 TeV. The detector consists of 86 strings of optical modules located within the Antarctic ice near the South Pole.   The energy and path-length through the Earth of these atmospheric neutrinos is equivalent to an $L/E \sim 1$ m/MeV value, similar to the SBL experiments.  However, the strength of the IceCube null result, shown in Fig.~\ref{MINOSIce}, arises from the additional 
modifications of the oscillation behavior when high energy neutrinos travel through dense matter, called ``matter effects.''

The matter-effect signature in IceCube corresponds to a predicted large deficit in the antineutrino flux for the up-going neutrinos that cross the Earth, given a 3+1 model with an anomalous squared mass splitting of $\sim 1$ eV$^2$ \cite{Nunokawa:2003ep,Choubey:2007ji, Razzaque:2012tp,Esmaili:2013vza,Barger:2011rc,Esmaili:2012nz,Esmaili:2013cja}.  This modification to the vacuum oscillation formalism comes from differences between neutrino
charged- and neutral-current interactions with the earth. In experiments at low energy or short baselines, this effect is negligible, and only vacuum oscillations need to be considered.   However, at the high energies and long baselines available to the IceCube experiment, coherent forward scattering can significantly affect neutrino propagation. 
In a 3+1 model, an additional matter potential is introduced to account for the difference of active flavor neutrinos scattering from matter--a contribution that is missing for the sterile flavor.

The matter effect is dependent on the neutrino mass hierarchy.
In the case of a ${3+1}$ hierarchy, as opposed to a ${1+3}$ hierarchy, the matter-induced resonance will appear in the antineutrino events rather than the neutrino events.  Detectable effects will lie in the range 0.01 $\leq \Delta m^2 \leq 10$~eV$^2$---the region of interest for our global fits. This follows from the the resonant energy: $E_{res} = \frac{\cos 2\theta \Delta m^2}{\sqrt{2} G_F N_{nuc}}$, where $\theta$ is an effective two flavor active-to-sterile neutrino mixing angle and $N_{nuc}$ is the target number density. The quoted sensitivity range can be understood by replacing $N_{nuc}$ by the corresponding density of the Earth, and the energy by the energy thresholds of the data set.  
It should be noted that the IceCube null result leads to a more restrictive limit in the case of 1+3 compared to a 3+1 model. This comes about because about 70\% of the events in IceCube are due to neutrino interactions, where a 1+3 signal would appear.   This is in agreement with the conclusions of cosmology and further justifies our concentration on 3+1 models below.

Use of matter effects in the IceCube analysis breaks degeneracies in the fits, allowing, for the first time, to constrain all of the elements of the 3+1 mixing matrix.
Examining Eq.~\ref{4mixmx},  one sees that the matrix has elements determined by the atmospheric and solar oscillation measurements, for which we use the results of Ref.~\cite{Parke:2015goa} as the range of allowed values.  This leaves seven further elements. 
Four of these elements,  ($U_{s1},\ldots,U_{s4}$), cannot be directly constrained by experiment due to the non-interacting nature of the `sterile' flavor state.    
However, in a 3+1 model, the mixing matrix is unitary, and
so these unmeasureable elements can be determined if the remaining three matrix elements are constrained. This leaves the elements $U_{e4}$, $U_{\mu4}$ and $U_{\tau 4}$ to be determined from the global fits to the combined SBL and IceCube data sets that we present below.

The SBL approximation, which has been applied in our previous fits \cite{Collin:2016rao}, cannot be applied when including the matter-effect signature in IceCube. In our global analysis, the $\nu$SM values of $7.5 \times 10^{-5} {\rm eV}^2$ and $2.3 \times 10^{-3} {\rm eV}^2$ from Ref.~\cite{nufit} are used for $\Delta m^2_{sol}$ and $\Delta m^2_{atm}$, respectively.  Furthermore, the introduction of IceCube data requires a parameterization of the extended lepton mixing matrix, Eq.~\ref{4mixmx}. This can be presented as a product of rotations following the convention specified in  Ref.~\cite{Delgado:2014kpa}:
\begin{equation}
U_{3+1} =
R_{34} R_{24} R_{14} R_{23} R_{13} R_{12}. \label{4mix_parametrization}
\end{equation}
Each $R_{ij}$ is a rotation matrix through angle $\theta_{ij}$ in the $ij$ plane. In this parameterization, the fourth column of $U_{3+1}$ is given by
\small\begin{align}
u_{4}=(&\sin\theta_{14},\thinspace\cos\theta_{14}\sin\theta_{24},\nonumber \\
& \thinspace\cos\theta_{14}\cos\theta_{24}\sin\theta_{34},\thinspace\cos\theta_{14}\cos\theta_{24}\cos\theta_{34})^{T}\,.\label{matrix_elements}
\end{align}\normalsize
If one sets all the CP violating phases to zero, only three new angles are introduced: $\theta_{14}$, $\theta_{24}$, and $\theta_{34}$. In addition, the IceCube collaboration analysis assumes $\theta_{14} = \theta_{34} = 0$.   
Under these assumptions, $\sin^22\theta_{24} = \sin^22\theta_{\mu \mu}$---the vacuum disappearance amplitude. 
While this is a convenient way to express the $\nu_\mu$ disappearance result (and is used in Ref.~\cite{IceCubePRL} along with other papers),
these assumptions will need to be relaxed in order to include IceCube in the global fits.

\begin{table*}[t]
\begin{center}
\begin{tabular}{|c|cccc|cccc|}\hline
3+1 & $\Delta m^2_{41}$ & $|U_{e 4}|$ & $|U_{\mu 4}|$ & $|U_{\tau 4}|$ & $N_{bins}$ & $\chi^2_{\mathrm{min}}$ & $\chi^2_{\mathrm{null}}$ & $\Delta \chi^2$ (dof)  \ \\  
\hline \hline
SBL  & 1.75 & 0.163 & 0.117 & - & 315 & 306.81 & 359.15  &52.34 (3)  \\  \hline
SBL+IC & 1.75 & 0.164  & 0.119 & 0.00  & 524 & 518.23 & 568.84 & 50.61 (4)\\ \hline
IC & 5.62 & - & 0.314 & - & 209 & 207.11 & 209.69 & 2.58 (2) \\ \hline
\end{tabular}
\end{center}

\caption{The oscillation parameter best-fit points for $3+1$ for the combined SBL and IceCube data sets compared to SBL alone. Units of $\Delta m^2$ are eV${}^2$.
\label{tab:bfpointsIce}}
\end{table*}

\begin{table}
\begin{center}
\begin{tabular}{|c|ccc|c|}\hline
$\Delta m^2/{\rm eV^2}$ & $|U_{e4}|$ & $|U_{\mu 4}|$ & $|U_{\tau 4}|$ & $\theta_{34}$ \\ \hline
6  & [0.17,0.21] & [0.10,0.13] & [0.00,0.05] & $ < 6\degree$ \\ \hline
2  & [0.13,0.20] & [0.09,0.15] & [0.00,0.70] & $ < 80\degree$ \\ \hline
\end{tabular}
\end{center}
\caption{The 90\% CL regions for matrix elements and the upper limit on $\theta_{34}$ for the two allowed regions in $\Delta m^2$.  For $\Delta m^2 = {\rm 1~ eV^2}$ there are no allowed regions at 90\%CL}
\label{tab:elementcls}
\end{table}

\begin{figure}[t!]
\center
\includegraphics[width=\columnwidth]{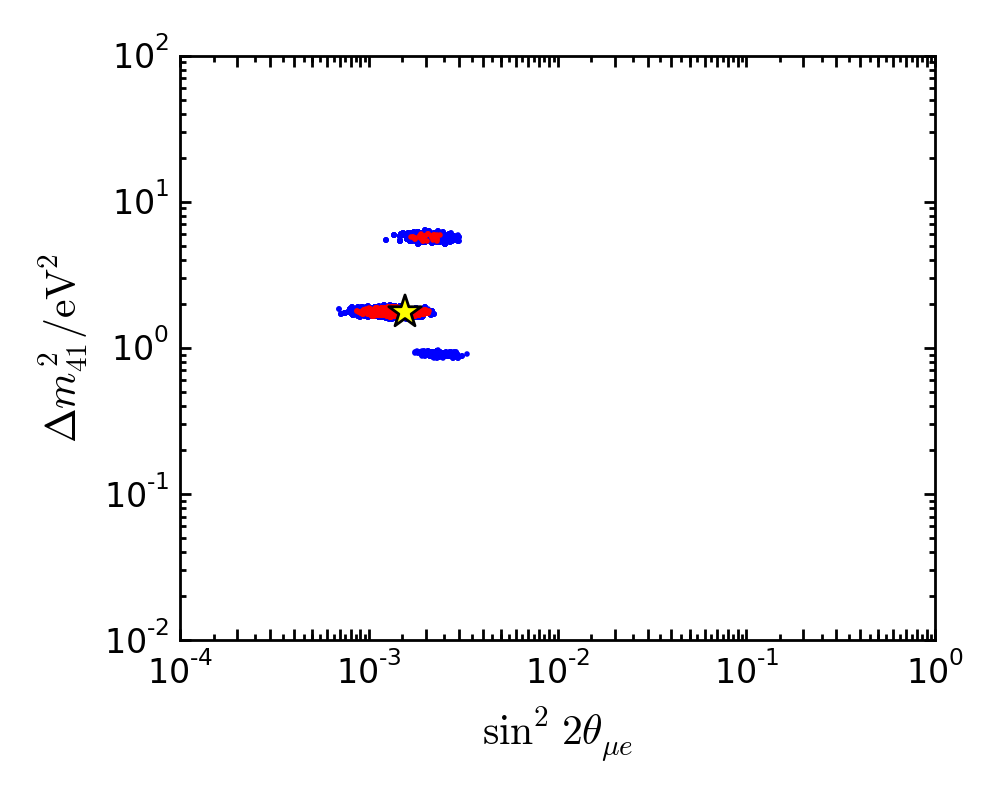}
\caption{Frequentist $3+1$ global fit for SBL+IceCube: $\Delta m^2_{4 1}$ vs $\sin^2{2\theta_{\mu e}}$. Red -- 90\% CL; Blue --99\% CL.  \label{fig:3plus1ice_sin22th}}
\end{figure}

\begin{figure}[t!]
\center
\includegraphics[width=\columnwidth]{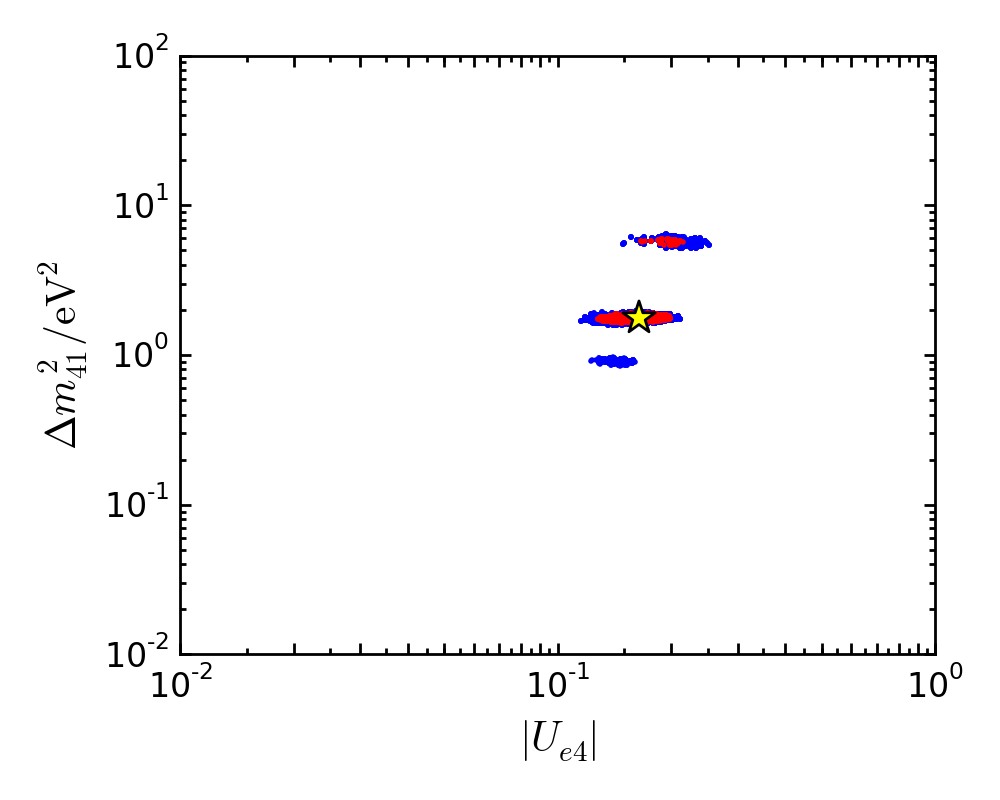} \\
\includegraphics[width=\columnwidth]{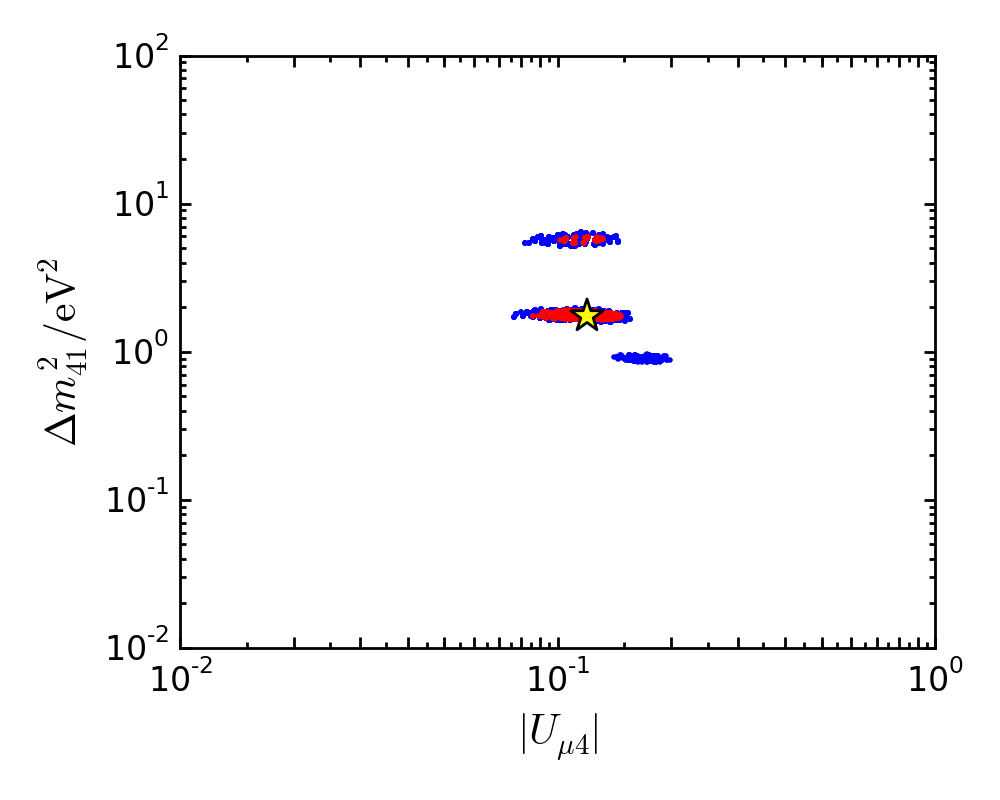}\\
\includegraphics[width=\columnwidth]{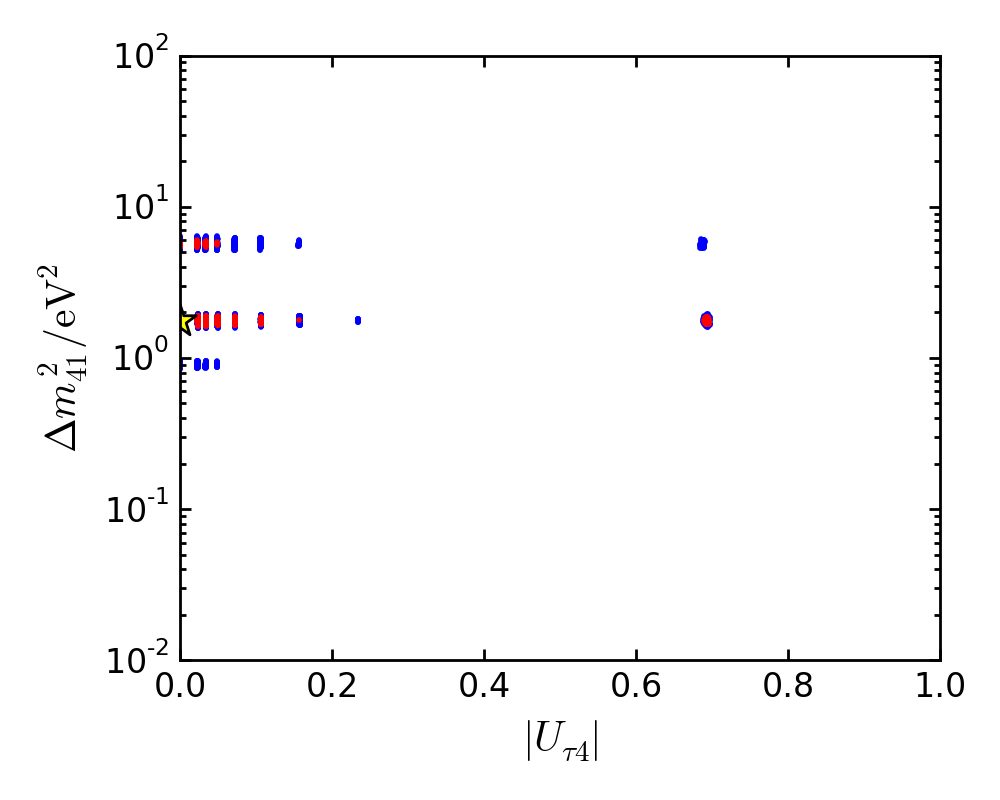}
\caption{Frequentist $3+1$ global fit SBL+IceCube, shown as a function of matrix element:  $|U_{e4}|$ (top), $|U_{\mu4}|$ (middle), and $|U_{\tau4}|$ (bottom).   Red -- 90\% CL; blue--99\% CL.\\ \label{fig:3plus1ice_U}}
\end{figure}

The IceCube analysis and the results presented here make use of the software package called nuSQuIDS \cite{Delgado:2014kpa,nusquids}. This software models flavor evolution from three ({\it i.e.} $\nu$SM) 
to six flavor basis states with customized matter potentials.   The 3+1 nuSQuIDS calculation does not use the short baseline approximation; thus, it includes the two additional $CP$ violating parameters that arise when $\Delta m^2_{21}$ and $\Delta m^2_{31}$ are nonzero and not equal.   However, in the case of the IceCube analysis, these $CP$ parameters are set to zero.  For the matter potential, nuSQuIDS makes use of the Preliminary Reference Earth Model (PREM) \cite{prem} parameterization describing the radial density profile of the Earth. The neutrino propagation implementation follows Eqs. (29-30) in Ref. \cite{GonzalezGarcia:2005xw}.  For the neutrino nucleon cross sections, we use the perturbative QCD calculation from Ref. \cite{pert,Cooper-Sarkar:2011fc}.

No evidence for anomalous $\nu_\mu$ or $\overline{\nu}_\mu$ disappearance was observed in the IceCube data set.  The resulting stringent limit extends to $\mathrm{sin}^2 2\theta_{24} \leq$  0.02 at $\Delta m^2 \sim$ 0.3 $\mathrm{eV}^2$ at 90\% CL for $\theta_{34}=0$ \cite{IceCubePRL}.
To incorporate this result into the fit, we must
relate the mixing angles $\theta_{14}$, $\theta_{24}$, and $\theta_{34}$ to the short-baseline neutrino oscillation probabilities. The oscillation amplitudes in this parameterization are found by substituting the matrix elements in Eq. \eqref{matrix_elements} into Eq. \eqref{PintermsofU}; e.g., $\sin^2{2\theta_{\mu e}} = \sin^2{2\theta_{1 4}} \sin^2{\theta_{2 4}}$. Since the short baseline anomalies imply $\sin^2{2\theta_{\mu e}} \neq 0$, it follows that we cannot assume
$\theta_{14}=0$ in a global fit.  

It has been shown \cite{Lindner:2015iaa} that  the presence of the matter-induced resonance critically depends on the value of $\theta_{34}$.  In particular, when $\theta_{34}$ is maximal, there is no matter-induced resonant enhancement. On the other hand, as noted by Ref. \cite{Esmaili:2013vza}, increasing $\theta_{34}$ distorts the atmospheric $\nu_\mu$ to $\nu_\tau$ neutrino oscillation. The interplay between these effects makes the IceCube data sensitive to $\theta_{34}$. We obtain the constraint on this parameter by sampling logarithmically in $\sin^2(2\theta_{34})$ from $10^{-3}$ to 1.   
The $CP$ phases have a sub-leading contribution in comparison to the $\theta_{34}$ effect \cite{Esmaili:2013vza};  thus, they have been set to zero. 

We describe the specific techniques of including the IceCube data into the fits in the appendix to this article.  Our capability of reproducing the IceCube result using the data release is shown in Fig.~\ref{MINOSIce}, dashed. The IceCube likelihood must be converted to a $\chi^2$ that can be combined with the SBL data.  The high computational cost of propagating neutrino fluxes through the Earth with nuSQuIDS prevents the analysis from being directly included into the global fitting software. Instead, the global fits were used to find a reduced set of parameters (``test points'') that could be evaluated directly. This assumes that the effect of IceCube on the global fit is a small perturbation, which is reasonable given that the IceCube-only $\Delta \chi^2$ is small compared to the SBL only global fit $\Delta \chi^2$ (see Table \ref{tab:bfpointsIce}).

\section{Results}

Figs.~\ref{fig:3plus1ice_sin22th} and \ref{fig:3plus1ice_U} show the SBL+IceCube global $3+1$ fit result.
The former shows $\Delta m^2_{4 1}$ vs $\sin^2{2\theta_{\mu e}}$, as defined in Eq.~\ref{sin22theq}.  The latter presents the result as a function of mixing matrix element.  
The $|U_{\tau4}|$ result is presented on a linear scale because one test point, the preferred solution, is $|U_{\tau4}|=0.$

The IceCube data excludes the solution at $\sim 1$ eV$^2$ at 90\% CL, although that solution persists at 99\% CL.  This has important implications for future sterile neutrino searches designed to address the 1 eV$^2$ allowed region.  
For example, given the peak energy of the BNB neutrino beam \cite{SBNproposal},
the position of the ICARUS T600 detector at Fermilab will result in a large potential signal for 1 eV$^2$ sterile neutrino, but less so if the $\Delta m^2$ is higher. 

As discussed, the SBL experiments constrain $|U_{e4}|$ and $|U_{\mu 4}|$, while the IceCube analysis has strong dependence on $|U_{\mu 4}|$ and $|U_{\tau 4}|$ through the matter-induced resonance. Thus, including IceCube provides insight into the less explored $|U_{\tau 4}|$ parameter.  Using $|U_{\tau 4}| = \cos\theta_{14}\cos\theta_{24}\sin\theta_{34}$, we convert the results to the 90\% C.L. ranges in Tab.~\ref{tab:elementcls}.   At $\Delta m^2 \sim 6$ eV$^2$, our limit  
improves the bound of $\theta_{34}<25\degree$ at 90\% C.L. from MINOS \cite{Adamson:2011ku} by a factor of four.

For the first time, this new result on $|U_{\tau 4}|$ allows us to have a complete picture of the extended lepton mixing matrix:\\ 
$|U|=$
\begin{align}    
 \scalemath{0.83}{\begin{bmatrix}
0.79 \rightarrow 0.83   \hspace{0.1cm} &0.53 \rightarrow 0.57 \hspace{0.1cm} & 0.14 \rightarrow 0.15 \hspace{0.1cm} & 0.13 \;(0.17) \rightarrow 0.20\:(0.21)\\
   0.25 \rightarrow 0.50 \hspace{0.1cm} & 0.46 \rightarrow 0.66 \hspace{0.1cm} & 0.64 \rightarrow 0.77 \hspace{0.1cm} & 0.09\;(0.10) \rightarrow 0.15\:(0.13)\\
  0.26 \rightarrow 0.54 \hspace{0.1cm}& 0.48 \rightarrow 0.69 \hspace{0.1cm} & 0.56 \rightarrow 0.75 \hspace{0.1cm} & 0.0\;(0.0)  \rightarrow 0.7\:(0.05)\\
     \ldots \hspace{0.1cm}& \ldots \hspace{0.1cm} & \ldots \hspace{0.1cm} & \ldots 
 \end{bmatrix}}.
\label{mix_matrix}\,
\end{align}
Above, ``\ldots''  represents parameters constrained by the overall unitarity of the $4\times4$ matrix.  The ranges in the matrix correspond to 90\% confidence intervals.
The entries in the last column correspond to this work and are given for $\Delta m^2 \sim 2~{\rm eV}^2$ ($\Delta m^2 \sim 6~{\rm eV}^2$). The intervals shown in each entry for the standard $3\times3$ submatrix were obtained from Ref. \cite{sparke:pricom}, and are independent of our fit. As a check of consistency, our values in the fourth column can be compared with the upper bounds from the $3\times3$ non-unitarity analysis in Ref. \cite{sparke:pricom}, which gave $|U_{e 4}| < 0.27$, $|U_{\mu 4}| < 0.73$, and $|U_{\tau 4}| < 0.623$ at 90\%CL. Our results in Eq.~\ref{mix_matrix} are fully compatible with these upper limits, which are based on standard 3-neutrino oscillation measurements exclusive of any sterile neutrino search data. 

\section{Conclusion}

We have presented three new results.   First, we have presented a combined fit of SBL and IceCube data resulting in a best fit of $\Delta m^2_{41} = 1.75\; \text{eV}^2$ with $\Delta \chi^2_{null-min}$ of 50.61 for 4 dof, which is not significantly different from past recent fits \cite{Collin:2016rao}.   Instead, the significance of the IceCube data is that it substantially lessens the likelihood of the $\sim 1$ eV$^2$ allowed region that was, until recently, the best fit point \cite{Conrad:2012qt}, and which has formed the basis of the design of many planned sterile neutrino searches.  Thus, despite not changing the best fit point, IceCube has significantly altered the sterile neutrino landscape.
Second, we have shown that this fit is sensitive to  $|U_{\tau 4}|$, providing improved constraint on $\theta_{34}$ of $< 80^\circ (< 6^\circ) $ at 90\% C.L. for $\Delta m^2_{41} \approx 2 (6)~\text{eV}^2$.   Lastly, we have used this, along with constraints from fits to atmospheric and solar data sets, to fill in all components of the 3+1 mixing matrix for the first time.


\section*{Acknowledgements}
GC, CA and JC are supported by NSF grants 1505858 and 1505855, and MS is  supported  by  NSF  grant  1404209. We thank Kevork Abazajian, Christina Ignarra, Benjamin Jones, William Louis, and Jordi Salvado for useful discussion.  We thank Maxim Goncharov for computing support.

\newpage

\bibliographystyle{apsrev}
\bibliography{nus_fit}

\begin{thebibliography}{72}
\expandafter\ifx\csname natexlab\endcsname\relax\def\natexlab#1{#1}\fi
\expandafter\ifx\csname bibnamefont\endcsname\relax
  \def\bibnamefont#1{#1}\fi
\expandafter\ifx\csname bibfnamefont\endcsname\relax
  \def\bibfnamefont#1{#1}\fi
\expandafter\ifx\csname citenamefont\endcsname\relax
  \def\citenamefont#1{#1}\fi
\expandafter\ifx\csname url\endcsname\relax
  \def\url#1{\texttt{#1}}\fi
\expandafter\ifx\csname urlprefix\endcsname\relax\def\urlprefix{URL }\fi
\providecommand{\bibinfo}[2]{#2}
\providecommand{\eprint}[2][]{\url{#2}}

\bibitem[{\citenamefont{Olive et~al.}(2014)}]{pdg}
\bibinfo{author}{\bibfnamefont{K.~A.} \bibnamefont{Olive}} \bibnamefont{et~al.}
  (\bibinfo{collaboration}{Particle Data Group}), \bibinfo{journal}{Chin.
  Phys.} \textbf{\bibinfo{volume}{C38}}, \bibinfo{pages}{090001}
  (\bibinfo{year}{2014}).

\bibitem[{\citenamefont{Gonzalez-Garcia
  et~al.}(2014)\citenamefont{Gonzalez-Garcia, Maltoni, and Schwetz}}]{nufit}
\bibinfo{author}{\bibfnamefont{M.}~\bibnamefont{Gonzalez-Garcia}},
  \bibinfo{author}{\bibfnamefont{M.}~\bibnamefont{Maltoni}}, \bibnamefont{and}
  \bibinfo{author}{\bibfnamefont{T.}~\bibnamefont{Schwetz}},
  \bibinfo{journal}{Journal of High Energy Physics}
  \textbf{\bibinfo{volume}{2014}}, \bibinfo{eid}{52} (\bibinfo{year}{2014}),
  \urlprefix\url{http://dx.doi.org/10.1007/JHEP11%282014%29052}.

\bibitem[{\citenamefont{Aguilar-Arevalo et~al.}(2001)}]{LSND}
\bibinfo{author}{\bibfnamefont{A.}~\bibnamefont{Aguilar-Arevalo}}
  \bibnamefont{et~al.} (\bibinfo{collaboration}{LSND Collaboration}),
  \bibinfo{journal}{Phys. Rev. D} \textbf{\bibinfo{volume}{64}},
  \bibinfo{pages}{112007} (\bibinfo{year}{2001}).

\bibitem[{\citenamefont{Aguilar-Arevalo et~al.}(2013)}]{nubarminiosc2}
\bibinfo{author}{\bibfnamefont{A.~A.} \bibnamefont{Aguilar-Arevalo}}
  \bibnamefont{et~al.} (\bibinfo{collaboration}{MiniBooNE}),
  \bibinfo{journal}{Phys. Rev. Lett.} \textbf{\bibinfo{volume}{110}},
  \bibinfo{pages}{161801} (\bibinfo{year}{2013}), \eprint{1207.4809}.

\bibitem[{\citenamefont{Aguilar-Arevalo
  et~al.}(2009{\natexlab{a}})}]{miniboonelowe2}
\bibinfo{author}{\bibfnamefont{A.~A.} \bibnamefont{Aguilar-Arevalo}}
  \bibnamefont{et~al.} (\bibinfo{collaboration}{MiniBooNE}),
  \bibinfo{journal}{Phys. Rev. Lett.} \textbf{\bibinfo{volume}{102}},
  \bibinfo{pages}{101802} (\bibinfo{year}{2009}{\natexlab{a}}),
  \eprint{0812.2243}.

\bibitem[{\citenamefont{Declais et~al.}(1995)\citenamefont{Declais, Favier,
  Metref, Pessard, Achkar et~al.}}]{Bugey}
\bibinfo{author}{\bibfnamefont{Y.}~\bibnamefont{Declais}},
  \bibinfo{author}{\bibfnamefont{J.}~\bibnamefont{Favier}},
  \bibinfo{author}{\bibfnamefont{A.}~\bibnamefont{Metref}},
  \bibinfo{author}{\bibfnamefont{H.}~\bibnamefont{Pessard}},
  \bibinfo{author}{\bibfnamefont{B.}~\bibnamefont{Achkar}},
  \bibnamefont{et~al.}, \bibinfo{journal}{Nucl. Phys. B}
  \textbf{\bibinfo{volume}{434}}, \bibinfo{pages}{503} (\bibinfo{year}{1995}).

\bibitem[{\citenamefont{Mention et~al.}(2011)\citenamefont{Mention, Fechner,
  Lasserre, Mueller, Lhuillier, Cribier, and Letourneau}}]{mention}
\bibinfo{author}{\bibfnamefont{G.}~\bibnamefont{Mention}},
  \bibinfo{author}{\bibfnamefont{M.}~\bibnamefont{Fechner}},
  \bibinfo{author}{\bibfnamefont{T.}~\bibnamefont{Lasserre}},
  \bibinfo{author}{\bibfnamefont{T.~A.} \bibnamefont{Mueller}},
  \bibinfo{author}{\bibfnamefont{D.}~\bibnamefont{Lhuillier}},
  \bibinfo{author}{\bibfnamefont{M.}~\bibnamefont{Cribier}}, \bibnamefont{and}
  \bibinfo{author}{\bibfnamefont{A.}~\bibnamefont{Letourneau}},
  \bibinfo{journal}{Phys. Rev.} \textbf{\bibinfo{volume}{D83}},
  \bibinfo{pages}{073006} (\bibinfo{year}{2011}), \eprint{1101.2755}.

\bibitem[{\citenamefont{Abdurashitov et~al.}(2009)}]{SAGE3}
\bibinfo{author}{\bibfnamefont{J.}~\bibnamefont{Abdurashitov}}
  \bibnamefont{et~al.} (\bibinfo{collaboration}{SAGE Collaboration}),
  \bibinfo{journal}{Phys. Rev. C} \textbf{\bibinfo{volume}{80}},
  \bibinfo{pages}{015807} (\bibinfo{year}{2009}).

\bibitem[{\citenamefont{Kaether et~al.}(2010)\citenamefont{Kaether, Hampel,
  Heusser, Kiko, and Kirsten}}]{GALLEX3}
\bibinfo{author}{\bibfnamefont{F.}~\bibnamefont{Kaether}},
  \bibinfo{author}{\bibfnamefont{W.}~\bibnamefont{Hampel}},
  \bibinfo{author}{\bibfnamefont{G.}~\bibnamefont{Heusser}},
  \bibinfo{author}{\bibfnamefont{J.}~\bibnamefont{Kiko}}, \bibnamefont{and}
  \bibinfo{author}{\bibfnamefont{T.}~\bibnamefont{Kirsten}},
  \bibinfo{journal}{Phys. Lett. B} \textbf{\bibinfo{volume}{685}},
  \bibinfo{pages}{47} (\bibinfo{year}{2010}).

\bibitem[{\citenamefont{{ALEPH Collaboration, DELPHI Collaboration, L3
  Collaboration, OPAL Collaboration, SLD Collaboration, LEP Electroweak Working
  Group, SLD Electroweak and Heavy Flavour Groups}}(2006)}]{zwidth}
\bibinfo{author}{\bibnamefont{{ALEPH Collaboration, DELPHI Collaboration, L3
  Collaboration, OPAL Collaboration, SLD Collaboration, LEP Electroweak Working
  Group, SLD Electroweak and Heavy Flavour Groups}}}, \bibinfo{journal}{Phys.
  Rept.} \textbf{\bibinfo{volume}{427}}, \bibinfo{pages}{257}
  (\bibinfo{year}{2006}).

\bibitem[{\citenamefont{Armbruster et~al.}(2002)}]{Karmen}
\bibinfo{author}{\bibfnamefont{B.}~\bibnamefont{Armbruster}}
  \bibnamefont{et~al.} (\bibinfo{collaboration}{KARMEN Collaboration}),
  \bibinfo{journal}{Phys. Rev. D} \textbf{\bibinfo{volume}{65}},
  \bibinfo{pages}{112001} (\bibinfo{year}{2002}).

\bibitem[{\citenamefont{Conrad and Shaevitz}(2012)}]{ConradShaevitz}
\bibinfo{author}{\bibfnamefont{J.}~\bibnamefont{Conrad}} \bibnamefont{and}
  \bibinfo{author}{\bibfnamefont{M.}~\bibnamefont{Shaevitz}},
  \bibinfo{journal}{Phys. Rev. D} \textbf{\bibinfo{volume}{85}},
  \bibinfo{pages}{013017} (\bibinfo{year}{2012}).

\bibitem[{\citenamefont{Adamson et~al.}(2009{\natexlab{a}})}]{MBNuMI}
\bibinfo{author}{\bibfnamefont{P.}~\bibnamefont{Adamson}} \bibnamefont{et~al.}
  (\bibinfo{collaboration}{MiniBooNE and MINOS Collaborations}),
  \bibinfo{journal}{Phys. Rev. Lett.} \textbf{\bibinfo{volume}{102}},
  \bibinfo{pages}{211801} (\bibinfo{year}{2009}{\natexlab{a}}).

\bibitem[{\citenamefont{Aguilar-Arevalo et~al.}(2009{\natexlab{b}})}]{Kendall}
\bibinfo{author}{\bibfnamefont{A.~A.} \bibnamefont{Aguilar-Arevalo}}
  \bibnamefont{et~al.} (\bibinfo{collaboration}{MiniBooNE Collaboration}),
  \bibinfo{journal}{Phys. Rev. Lett.} \textbf{\bibinfo{volume}{103}},
  \bibinfo{pages}{061802} (\bibinfo{year}{2009}{\natexlab{b}}).

\bibitem[{\citenamefont{Cheng et~al.}(2011)}]{K1}
\bibinfo{author}{\bibfnamefont{G.}~\bibnamefont{Cheng}} \bibnamefont{et~al.}
  (\bibinfo{collaboration}{SciBooNE Collaboration}), \bibinfo{journal}{Phys.
  Rev. D} \textbf{\bibinfo{volume}{84}}, \bibinfo{pages}{012009}
  (\bibinfo{year}{2011}).

\bibitem[{\citenamefont{Astier et~al.}(2003)}]{NOMAD1}
\bibinfo{author}{\bibfnamefont{P.}~\bibnamefont{Astier}} \bibnamefont{et~al.}
  (\bibinfo{collaboration}{NOMAD Collaboration}), \bibinfo{journal}{Phys. Lett.
  B} \textbf{\bibinfo{volume}{570}}, \bibinfo{pages}{19}
  (\bibinfo{year}{2003}).

\bibitem[{\citenamefont{Stockdale et~al.}(1985)\citenamefont{Stockdale, Bodek,
  Borcherding, Giokaris, Lang et~al.}}]{CCFR84}
\bibinfo{author}{\bibfnamefont{I.}~\bibnamefont{Stockdale}},
  \bibinfo{author}{\bibfnamefont{A.}~\bibnamefont{Bodek}},
  \bibinfo{author}{\bibfnamefont{F.}~\bibnamefont{Borcherding}},
  \bibinfo{author}{\bibfnamefont{N.}~\bibnamefont{Giokaris}},
  \bibinfo{author}{\bibfnamefont{K.}~\bibnamefont{Lang}}, \bibnamefont{et~al.},
  \bibinfo{journal}{Z. Phys. C} \textbf{\bibinfo{volume}{27}},
  \bibinfo{pages}{53} (\bibinfo{year}{1985}).

\bibitem[{\citenamefont{Dydak et~al.}(1984)\citenamefont{Dydak, Feldman, Guyot,
  Merlo, Meyer et~al.}}]{CDHS}
\bibinfo{author}{\bibfnamefont{F.}~\bibnamefont{Dydak}},
  \bibinfo{author}{\bibfnamefont{G.}~\bibnamefont{Feldman}},
  \bibinfo{author}{\bibfnamefont{C.}~\bibnamefont{Guyot}},
  \bibinfo{author}{\bibfnamefont{J.}~\bibnamefont{Merlo}},
  \bibinfo{author}{\bibfnamefont{H.}~\bibnamefont{Meyer}},
  \bibnamefont{et~al.}, \bibinfo{journal}{Phys. Lett. B}
  \textbf{\bibinfo{volume}{134}}, \bibinfo{pages}{281} (\bibinfo{year}{1984}).

\bibitem[{\citenamefont{{D.G. Michael {\it et al.} (MINOS
  Collaboration)}}(2006)}]{MINOSCC1}
\bibinfo{author}{\bibnamefont{{D.G. Michael {\it et al.} (MINOS
  Collaboration)}}}, \bibinfo{journal}{Phys. Rev. Lett.}
  \textbf{\bibinfo{volume}{97}}, \bibinfo{pages}{191801}
  (\bibinfo{year}{2006}).

\bibitem[{\citenamefont{{P. Adamson {\it et al.} (MINOS
  Collaboration)}}(2008{\natexlab{a}})}]{MINOSCC3}
\bibinfo{author}{\bibnamefont{{P. Adamson {\it et al.} (MINOS
  Collaboration)}}}, \bibinfo{journal}{Phys. Rev. Lett.}
  \textbf{\bibinfo{volume}{101}}, \bibinfo{pages}{131802}
  (\bibinfo{year}{2008}{\natexlab{a}}).

\bibitem[{\citenamefont{Collin et~al.}(2016)\citenamefont{Collin, Argüelles,
  Conrad, and Shaevitz}}]{Collin:2016rao}
\bibinfo{author}{\bibfnamefont{G.~H.} \bibnamefont{Collin}},
  \bibinfo{author}{\bibfnamefont{C.~A.} \bibnamefont{Argüelles}},
  \bibinfo{author}{\bibfnamefont{J.~M.} \bibnamefont{Conrad}},
  \bibnamefont{and} \bibinfo{author}{\bibfnamefont{M.~H.}
  \bibnamefont{Shaevitz}}, \bibinfo{journal}{Nucl. Phys.}
  \textbf{\bibinfo{volume}{B908}}, \bibinfo{pages}{354} (\bibinfo{year}{2016}),
  \eprint{1602.00671}.

\bibitem[{\citenamefont{Kopp et~al.}(2013)\citenamefont{Kopp, Machado, Maltoni,
  and Schwetz}}]{kopp_sterile_2013}
\bibinfo{author}{\bibfnamefont{J.}~\bibnamefont{Kopp}},
  \bibinfo{author}{\bibfnamefont{P.~A.~N.} \bibnamefont{Machado}},
  \bibinfo{author}{\bibfnamefont{M.}~\bibnamefont{Maltoni}}, \bibnamefont{and}
  \bibinfo{author}{\bibfnamefont{T.}~\bibnamefont{Schwetz}},
  \bibinfo{journal}{Journal of High Energy Physics}
  \textbf{\bibinfo{volume}{2013}} (\bibinfo{year}{2013}), ISSN
  \bibinfo{issn}{1029-8479}, \bibinfo{note}{arXiv: 1303.3011},
  \urlprefix\url{http://arxiv.org/abs/1303.3011}.

\bibitem[{\citenamefont{Giunti et~al.}(2013)\citenamefont{Giunti, Laveder, Li,
  and Long}}]{giunti_pragmatic_2013}
\bibinfo{author}{\bibfnamefont{C.}~\bibnamefont{Giunti}},
  \bibinfo{author}{\bibfnamefont{M.}~\bibnamefont{Laveder}},
  \bibinfo{author}{\bibfnamefont{Y.~F.} \bibnamefont{Li}}, \bibnamefont{and}
  \bibinfo{author}{\bibfnamefont{H.~W.} \bibnamefont{Long}},
  \bibinfo{journal}{Physical Review D} \textbf{\bibinfo{volume}{88}}
  (\bibinfo{year}{2013}), ISSN \bibinfo{issn}{1550-7998, 1550-2368},
  \bibinfo{note}{arXiv: 1308.5288},
  \urlprefix\url{http://arxiv.org/abs/1308.5288}.

\bibitem[{\citenamefont{{H. Chen \textit{et al.} (MicroBooNE
  Collaboration)}}(2007)}]{uboone}
\bibinfo{author}{\bibnamefont{{H. Chen \textit{et al.} (MicroBooNE
  Collaboration)}}} (\bibinfo{year}{2007}), \eprint{{``Proposal for a New
  Experiment Using the Booster and NuMI Neutrino Beamlines: MicroBooNE''}},
  \urlprefix\url{http://www-microboone.fnal.gov/public/MicroBooNE_10152007.pdf}.

\bibitem[{\citenamefont{Antonello et~al.}(2015)}]{SBNproposal}
\bibinfo{author}{\bibfnamefont{M.}~\bibnamefont{Antonello}}
  \bibnamefont{et~al.} (\bibinfo{collaboration}{LAr1-ND, ICARUS-WA104,
  MicroBooNE}) (\bibinfo{year}{2015}), \eprint{1503.01520}.

\bibitem[{\citenamefont{Adelmann et~al.}(2014)\citenamefont{Adelmann, Alonso,
  Barletta, Conrad, Shaevitz, Spitz, Toups, and Winslow}}]{IsoDAR}
\bibinfo{author}{\bibfnamefont{A.}~\bibnamefont{Adelmann}},
  \bibinfo{author}{\bibfnamefont{J.}~\bibnamefont{Alonso}},
  \bibinfo{author}{\bibfnamefont{W.~A.} \bibnamefont{Barletta}},
  \bibinfo{author}{\bibfnamefont{J.~M.} \bibnamefont{Conrad}},
  \bibinfo{author}{\bibfnamefont{M.~H.} \bibnamefont{Shaevitz}},
  \bibinfo{author}{\bibfnamefont{J.}~\bibnamefont{Spitz}},
  \bibinfo{author}{\bibfnamefont{M.}~\bibnamefont{Toups}}, \bibnamefont{and}
  \bibinfo{author}{\bibfnamefont{L.~A.} \bibnamefont{Winslow}},
  \bibinfo{journal}{Adv. High Energy Phys.} \textbf{\bibinfo{volume}{2014}},
  \bibinfo{pages}{347097} (\bibinfo{year}{2014}), \eprint{1307.6465}.

\bibitem[{\citenamefont{Langford}(2015)}]{Prospect}
\bibinfo{author}{\bibfnamefont{T.~J.} \bibnamefont{Langford}}
  (\bibinfo{collaboration}{PROSPECT}), \bibinfo{journal}{Nucl. Part. Phys.
  Proc.} \textbf{\bibinfo{volume}{265-266}}, \bibinfo{pages}{123}
  (\bibinfo{year}{2015}), \eprint{1501.00194}.

\bibitem[{\citenamefont{Bravo-Berguño et~al.}(2016)}]{sox}
\bibinfo{author}{\bibfnamefont{D.}~\bibnamefont{Bravo-Berguño}}
  \bibnamefont{et~al.} (\bibinfo{collaboration}{SOX}), \bibinfo{journal}{Nucl.
  Part. Phys. Proc.} \textbf{\bibinfo{volume}{273-275}}, \bibinfo{pages}{1760}
  (\bibinfo{year}{2016}).

\bibitem[{\citenamefont{Aartsen et~al.}(2016)}]{IceCubePRL}
\bibinfo{author}{\bibfnamefont{M.~G.} \bibnamefont{Aartsen}}
  \bibnamefont{et~al.} (\bibinfo{collaboration}{IceCube})
  (\bibinfo{year}{2016}), \eprint{1605.01990}.

\bibitem[{\citenamefont{Aguilar-Arevalo et~al.}(2007)}]{miniboonelowe1}
\bibinfo{author}{\bibfnamefont{A.~A.} \bibnamefont{Aguilar-Arevalo}}
  \bibnamefont{et~al.} (\bibinfo{collaboration}{MiniBooNE}),
  \bibinfo{journal}{Phys. Rev. Lett.} \textbf{\bibinfo{volume}{98}},
  \bibinfo{pages}{231801} (\bibinfo{year}{2007}), \eprint{0704.1500}.

\bibitem[{\citenamefont{Aguilar-Arevalo et~al.}(2010)}]{nubarminiosc1}
\bibinfo{author}{\bibfnamefont{A.~A.} \bibnamefont{Aguilar-Arevalo}}
  \bibnamefont{et~al.} (\bibinfo{collaboration}{MiniBooNE}),
  \bibinfo{journal}{Phys. Rev. Lett.} \textbf{\bibinfo{volume}{105}},
  \bibinfo{pages}{181801} (\bibinfo{year}{2010}), \eprint{1007.1150}.

\bibitem[{\citenamefont{Adamson et~al.}(2009{\natexlab{b}})}]{NuMIMB}
\bibinfo{author}{\bibfnamefont{P.}~\bibnamefont{Adamson}} \bibnamefont{et~al.}
  (\bibinfo{collaboration}{MiniBooNE, MINOS}), \bibinfo{journal}{Phys. Rev.
  Lett.} \textbf{\bibinfo{volume}{102}}, \bibinfo{pages}{211801}
  (\bibinfo{year}{2009}{\natexlab{b}}), \eprint{0809.2447}.

\bibitem[{\citenamefont{Mahn et~al.}(2012)}]{Mahn:2011ea}
\bibinfo{author}{\bibfnamefont{K.~B.~M.} \bibnamefont{Mahn}}
  \bibnamefont{et~al.} (\bibinfo{collaboration}{SciBooNE, MiniBooNE}),
  \bibinfo{journal}{Phys. Rev.} \textbf{\bibinfo{volume}{D85}},
  \bibinfo{pages}{032007} (\bibinfo{year}{2012}), \eprint{1106.5685}.

\bibitem[{\citenamefont{Cheng et~al.}(2012)}]{Cheng:2012yy}
\bibinfo{author}{\bibfnamefont{G.}~\bibnamefont{Cheng}} \bibnamefont{et~al.}
  (\bibinfo{collaboration}{SciBooNE, MiniBooNE}), \bibinfo{journal}{Phys. Rev.}
  \textbf{\bibinfo{volume}{D86}}, \bibinfo{pages}{052009}
  (\bibinfo{year}{2012}), \eprint{1208.0322}.

\bibitem[{\citenamefont{{P. Adamson {\it et al.} (MINOS
  Collaboration)}}(2008{\natexlab{b}})}]{MINOSCC2}
\bibinfo{author}{\bibnamefont{{P. Adamson {\it et al.} (MINOS
  Collaboration)}}}, \bibinfo{journal}{Phys. Rev. D}
  \textbf{\bibinfo{volume}{77}}, \bibinfo{pages}{072002}
  (\bibinfo{year}{2008}{\natexlab{b}}).

\bibitem[{\citenamefont{An et~al.}(2016)}]{An:2016luf}
\bibinfo{author}{\bibfnamefont{F.~P.} \bibnamefont{An}} \bibnamefont{et~al.}
  (\bibinfo{collaboration}{Daya Bay}) (\bibinfo{year}{2016}),
  \eprint{1607.01174}.

\bibitem[{\citenamefont{Adamson et~al.}(2016)}]{MINOS2016}
\bibinfo{author}{\bibfnamefont{P.}~\bibnamefont{Adamson}} \bibnamefont{et~al.}
  (\bibinfo{collaboration}{MINOS}), \bibinfo{journal}{Submitted to: Phys. Rev.
  Lett.}  (\bibinfo{year}{2016}), \eprint{1607.01176}.

\bibitem[{\citenamefont{Adamson et~al.}(2011{\natexlab{a}})}]{Minosth13}
\bibinfo{author}{\bibfnamefont{P.}~\bibnamefont{Adamson}} \bibnamefont{et~al.}
  (\bibinfo{collaboration}{MINOS}), \bibinfo{journal}{Phys. Rev. Lett.}
  \textbf{\bibinfo{volume}{107}}, \bibinfo{pages}{181802}
  (\bibinfo{year}{2011}{\natexlab{a}}), \eprint{1108.0015}.

\bibitem[{\citenamefont{Abazajian and Kaplinghat}(2016)}]{KevAnnRev}
\bibinfo{author}{\bibfnamefont{K.~N.} \bibnamefont{Abazajian}}
  \bibnamefont{and}
  \bibinfo{author}{\bibfnamefont{M.}~\bibnamefont{Kaplinghat}},
  \bibinfo{journal}{Ann Rev. Nuc. Part. Phys.} \textbf{\bibinfo{volume}{66}},
  \bibinfo{pages}{401} (\bibinfo{year}{2016}).

\bibitem[{\citenamefont{Abazajian}(2003)}]{Kev2}
\bibinfo{author}{\bibfnamefont{K.~N.} \bibnamefont{Abazajian}},
  \bibinfo{journal}{Astropart. Phys.} \textbf{\bibinfo{volume}{19}},
  \bibinfo{pages}{303} (\bibinfo{year}{2003}), \eprint{astro-ph/0205238}.

\bibitem[{\citenamefont{Dasgupta and Kopp}(2014)}]{Dasgupta:2013zpn}
\bibinfo{author}{\bibfnamefont{B.}~\bibnamefont{Dasgupta}} \bibnamefont{and}
  \bibinfo{author}{\bibfnamefont{J.}~\bibnamefont{Kopp}},
  \bibinfo{journal}{Phys. Rev. Lett.} \textbf{\bibinfo{volume}{112}},
  \bibinfo{pages}{031803} (\bibinfo{year}{2014}), \eprint{1310.6337}.

\bibitem[{\citenamefont{Hannestad et~al.}(2014)\citenamefont{Hannestad, Hansen,
  and Tram}}]{Hannestad:2013ana}
\bibinfo{author}{\bibfnamefont{S.}~\bibnamefont{Hannestad}},
  \bibinfo{author}{\bibfnamefont{R.~S.} \bibnamefont{Hansen}},
  \bibnamefont{and} \bibinfo{author}{\bibfnamefont{T.}~\bibnamefont{Tram}},
  \bibinfo{journal}{Phys. Rev. Lett.} \textbf{\bibinfo{volume}{112}},
  \bibinfo{pages}{031802} (\bibinfo{year}{2014}), \eprint{1310.5926}.

\bibitem[{\citenamefont{Bento and Berezhiani}(2001)}]{Bento:2001xi}
\bibinfo{author}{\bibfnamefont{L.}~\bibnamefont{Bento}} \bibnamefont{and}
  \bibinfo{author}{\bibfnamefont{Z.}~\bibnamefont{Berezhiani}},
  \bibinfo{journal}{Phys. Rev.} \textbf{\bibinfo{volume}{D64}},
  \bibinfo{pages}{115015} (\bibinfo{year}{2001}), \eprint{hep-ph/0108064}.

\bibitem[{\citenamefont{Chu and Cirelli}(2006)}]{Chu:2006ua}
\bibinfo{author}{\bibfnamefont{Y.-Z.} \bibnamefont{Chu}} \bibnamefont{and}
  \bibinfo{author}{\bibfnamefont{M.}~\bibnamefont{Cirelli}},
  \bibinfo{journal}{Phys. Rev.} \textbf{\bibinfo{volume}{D74}},
  \bibinfo{pages}{085015} (\bibinfo{year}{2006}), \eprint{astro-ph/0608206}.

\bibitem[{\citenamefont{Foot and Volkas}(1995)}]{Foot:1995bm}
\bibinfo{author}{\bibfnamefont{R.}~\bibnamefont{Foot}} \bibnamefont{and}
  \bibinfo{author}{\bibfnamefont{R.~R.} \bibnamefont{Volkas}},
  \bibinfo{journal}{Phys. Rev. Lett.} \textbf{\bibinfo{volume}{75}},
  \bibinfo{pages}{4350} (\bibinfo{year}{1995}), \eprint{hep-ph/9508275}.

\bibitem[{\citenamefont{Gelmini et~al.}(2004)\citenamefont{Gelmini,
  Palomares-Ruiz, and Pascoli}}]{Gelmini:2004ah}
\bibinfo{author}{\bibfnamefont{G.}~\bibnamefont{Gelmini}},
  \bibinfo{author}{\bibfnamefont{S.}~\bibnamefont{Palomares-Ruiz}},
  \bibnamefont{and} \bibinfo{author}{\bibfnamefont{S.}~\bibnamefont{Pascoli}},
  \bibinfo{journal}{Phys. Rev. Lett.} \textbf{\bibinfo{volume}{93}},
  \bibinfo{pages}{081302} (\bibinfo{year}{2004}), \eprint{astro-ph/0403323}.

\bibitem[{\citenamefont{Ho and Scherrer}(2013)}]{Ho:2012br}
\bibinfo{author}{\bibfnamefont{C.~M.} \bibnamefont{Ho}} \bibnamefont{and}
  \bibinfo{author}{\bibfnamefont{R.~J.} \bibnamefont{Scherrer}},
  \bibinfo{journal}{Phys. Rev.} \textbf{\bibinfo{volume}{D87}},
  \bibinfo{pages}{065016} (\bibinfo{year}{2013}), \eprint{1212.1689}.

\bibitem[{\citenamefont{Hamann et~al.}(2011)\citenamefont{Hamann, Hannestad,
  Raffelt, and Wong}}]{Hamann:2011ge}
\bibinfo{author}{\bibfnamefont{J.}~\bibnamefont{Hamann}},
  \bibinfo{author}{\bibfnamefont{S.}~\bibnamefont{Hannestad}},
  \bibinfo{author}{\bibfnamefont{G.~G.} \bibnamefont{Raffelt}},
  \bibnamefont{and} \bibinfo{author}{\bibfnamefont{Y.~Y.~Y.}
  \bibnamefont{Wong}}, \bibinfo{journal}{JCAP} \textbf{\bibinfo{volume}{1109}},
  \bibinfo{pages}{034} (\bibinfo{year}{2011}), \eprint{1108.4136}.

\bibitem[{\citenamefont{Abazajian et~al.}(2005)\citenamefont{Abazajian, Bell,
  Fuller, and Wong}}]{Kev1}
\bibinfo{author}{\bibfnamefont{K.}~\bibnamefont{Abazajian}},
  \bibinfo{author}{\bibfnamefont{N.~F.} \bibnamefont{Bell}},
  \bibinfo{author}{\bibfnamefont{G.~M.} \bibnamefont{Fuller}},
  \bibnamefont{and} \bibinfo{author}{\bibfnamefont{Y.~Y.~Y.}
  \bibnamefont{Wong}}, \bibinfo{journal}{Phys. Rev.}
  \textbf{\bibinfo{volume}{D72}}, \bibinfo{pages}{063004}
  (\bibinfo{year}{2005}), \eprint{astro-ph/0410175}.

\bibitem[{\citenamefont{Archidiacono et~al.}(2016)\citenamefont{Archidiacono,
  Gariazzo, Giunti, Hannestad, Hansen, Laveder, and Tram}}]{Sten}
\bibinfo{author}{\bibfnamefont{M.}~\bibnamefont{Archidiacono}},
  \bibinfo{author}{\bibfnamefont{S.}~\bibnamefont{Gariazzo}},
  \bibinfo{author}{\bibfnamefont{C.}~\bibnamefont{Giunti}},
  \bibinfo{author}{\bibfnamefont{S.}~\bibnamefont{Hannestad}},
  \bibinfo{author}{\bibfnamefont{R.}~\bibnamefont{Hansen}},
  \bibinfo{author}{\bibfnamefont{M.}~\bibnamefont{Laveder}}, \bibnamefont{and}
  \bibinfo{author}{\bibfnamefont{T.}~\bibnamefont{Tram}},
  \bibinfo{journal}{JCAP} \textbf{\bibinfo{volume}{1608}}, \bibinfo{pages}{067}
  (\bibinfo{year}{2016}), \eprint{1606.07673}.

\bibitem[{\citenamefont{Nunokawa et~al.}(2003)\citenamefont{Nunokawa, Peres,
  and Zukanovich~Funchal}}]{Nunokawa:2003ep}
\bibinfo{author}{\bibfnamefont{H.}~\bibnamefont{Nunokawa}},
  \bibinfo{author}{\bibfnamefont{O.~L.~G.} \bibnamefont{Peres}},
  \bibnamefont{and}
  \bibinfo{author}{\bibfnamefont{R.}~\bibnamefont{Zukanovich~Funchal}},
  \bibinfo{journal}{Phys. Lett.} \textbf{\bibinfo{volume}{B562}},
  \bibinfo{pages}{279} (\bibinfo{year}{2003}), \eprint{hep-ph/0302039}.

\bibitem[{\citenamefont{Choubey}(2007)}]{Choubey:2007ji}
\bibinfo{author}{\bibfnamefont{S.}~\bibnamefont{Choubey}},
  \bibinfo{journal}{JHEP} \textbf{\bibinfo{volume}{12}}, \bibinfo{pages}{014}
  (\bibinfo{year}{2007}), \eprint{0709.1937}.

\bibitem[{\citenamefont{Razzaque and Smirnov}(2012)}]{Razzaque:2012tp}
\bibinfo{author}{\bibfnamefont{S.}~\bibnamefont{Razzaque}} \bibnamefont{and}
  \bibinfo{author}{\bibfnamefont{A.~{\relax Yu}.} \bibnamefont{Smirnov}},
  \bibinfo{journal}{Phys. Rev.} \textbf{\bibinfo{volume}{D85}},
  \bibinfo{pages}{093010} (\bibinfo{year}{2012}), \eprint{1203.5406}.

\bibitem[{\citenamefont{Esmaili and Smirnov}(2013)}]{Esmaili:2013vza}
\bibinfo{author}{\bibfnamefont{A.}~\bibnamefont{Esmaili}} \bibnamefont{and}
  \bibinfo{author}{\bibfnamefont{A.~{\relax Yu}.} \bibnamefont{Smirnov}},
  \bibinfo{journal}{JHEP} \textbf{\bibinfo{volume}{12}}, \bibinfo{pages}{014}
  (\bibinfo{year}{2013}), \eprint{1307.6824}.

\bibitem[{\citenamefont{Barger et~al.}(2012)\citenamefont{Barger, Gao, and
  Marfatia}}]{Barger:2011rc}
\bibinfo{author}{\bibfnamefont{V.}~\bibnamefont{Barger}},
  \bibinfo{author}{\bibfnamefont{Y.}~\bibnamefont{Gao}}, \bibnamefont{and}
  \bibinfo{author}{\bibfnamefont{D.}~\bibnamefont{Marfatia}},
  \bibinfo{journal}{Phys. Rev.} \textbf{\bibinfo{volume}{D85}},
  \bibinfo{pages}{011302} (\bibinfo{year}{2012}), \eprint{1109.5748}.

\bibitem[{\citenamefont{Esmaili et~al.}(2012)\citenamefont{Esmaili, Halzen, and
  Peres}}]{Esmaili:2012nz}
\bibinfo{author}{\bibfnamefont{A.}~\bibnamefont{Esmaili}},
  \bibinfo{author}{\bibfnamefont{F.}~\bibnamefont{Halzen}}, \bibnamefont{and}
  \bibinfo{author}{\bibfnamefont{O.~L.~G.} \bibnamefont{Peres}},
  \bibinfo{journal}{JCAP} \textbf{\bibinfo{volume}{1211}}, \bibinfo{pages}{041}
  (\bibinfo{year}{2012}), \eprint{1206.6903}.

\bibitem[{\citenamefont{Esmaili et~al.}(2013)\citenamefont{Esmaili, Halzen, and
  Peres}}]{Esmaili:2013cja}
\bibinfo{author}{\bibfnamefont{A.}~\bibnamefont{Esmaili}},
  \bibinfo{author}{\bibfnamefont{F.}~\bibnamefont{Halzen}}, \bibnamefont{and}
  \bibinfo{author}{\bibfnamefont{O.~L.~G.} \bibnamefont{Peres}},
  \bibinfo{journal}{JCAP} \textbf{\bibinfo{volume}{1307}}, \bibinfo{pages}{048}
  (\bibinfo{year}{2013}), \eprint{1303.3294}.

\bibitem[{\citenamefont{Parke and Ross-Lonergan}(2015)}]{Parke:2015goa}
\bibinfo{author}{\bibfnamefont{S.}~\bibnamefont{Parke}} \bibnamefont{and}
  \bibinfo{author}{\bibfnamefont{M.}~\bibnamefont{Ross-Lonergan}}
  (\bibinfo{year}{2015}), \eprint{1508.05095}.

\bibitem[{\citenamefont{Arguelles~Delgado
  et~al.}(2014)\citenamefont{Arguelles~Delgado, Salvado, and
  Weaver}}]{Delgado:2014kpa}
\bibinfo{author}{\bibfnamefont{C.~A.} \bibnamefont{Arguelles~Delgado}},
  \bibinfo{author}{\bibfnamefont{J.}~\bibnamefont{Salvado}}, \bibnamefont{and}
  \bibinfo{author}{\bibfnamefont{C.~N.} \bibnamefont{Weaver}}
  (\bibinfo{year}{2014}), \eprint{1412.3832}.

\bibitem[{\citenamefont{Arguelles~Delgado
  et~al.}()\citenamefont{Arguelles~Delgado, Salvado, and Weaver}}]{nusquids}
\bibinfo{author}{\bibfnamefont{C.~A.} \bibnamefont{Arguelles~Delgado}},
  \bibinfo{author}{\bibfnamefont{J.}~\bibnamefont{Salvado}}, \bibnamefont{and}
  \bibinfo{author}{\bibfnamefont{C.~N.} \bibnamefont{Weaver}},
  \emph{\bibinfo{title}{{$\nu$-SQuIDS}}},
  \bibinfo{howpublished}{\url{https://github.com/arguelles/nuSQuIDS}}.

\bibitem[{\citenamefont{{Dziewonski} and {Anderson}}(1981)}]{prem}
\bibinfo{author}{\bibfnamefont{A.~M.} \bibnamefont{{Dziewonski}}}
  \bibnamefont{and} \bibinfo{author}{\bibfnamefont{D.~L.}
  \bibnamefont{{Anderson}}}, \bibinfo{journal}{Physics of the Earth and
  Planetary Interiors} \textbf{\bibinfo{volume}{25}}, \bibinfo{pages}{297}
  (\bibinfo{year}{1981}).

\bibitem[{\citenamefont{Gonzalez-Garcia
  et~al.}(2005)\citenamefont{Gonzalez-Garcia, Halzen, and
  Maltoni}}]{GonzalezGarcia:2005xw}
\bibinfo{author}{\bibfnamefont{M.~C.} \bibnamefont{Gonzalez-Garcia}},
  \bibinfo{author}{\bibfnamefont{F.}~\bibnamefont{Halzen}}, \bibnamefont{and}
  \bibinfo{author}{\bibfnamefont{M.}~\bibnamefont{Maltoni}},
  \bibinfo{journal}{Phys. Rev.} \textbf{\bibinfo{volume}{D71}},
  \bibinfo{pages}{093010} (\bibinfo{year}{2005}), \eprint{hep-ph/0502223}.

\bibitem[{\citenamefont{Arguelles et~al.}(2015)\citenamefont{Arguelles, Halzen,
  Wille, Kroll, and Reno}}]{pert}
\bibinfo{author}{\bibfnamefont{C.~A.} \bibnamefont{Arguelles}},
  \bibinfo{author}{\bibfnamefont{F.}~\bibnamefont{Halzen}},
  \bibinfo{author}{\bibfnamefont{L.}~\bibnamefont{Wille}},
  \bibinfo{author}{\bibfnamefont{M.}~\bibnamefont{Kroll}}, \bibnamefont{and}
  \bibinfo{author}{\bibfnamefont{M.~H.} \bibnamefont{Reno}},
  \bibinfo{journal}{Phys. Rev.} \textbf{\bibinfo{volume}{D92}},
  \bibinfo{pages}{074040} (\bibinfo{year}{2015}), \eprint{1504.06639}.

\bibitem[{\citenamefont{Cooper-Sarkar et~al.}(2011)\citenamefont{Cooper-Sarkar,
  Mertsch, and Sarkar}}]{Cooper-Sarkar:2011fc}
\bibinfo{author}{\bibfnamefont{A.}~\bibnamefont{Cooper-Sarkar}},
  \bibinfo{author}{\bibfnamefont{P.}~\bibnamefont{Mertsch}}, \bibnamefont{and}
  \bibinfo{author}{\bibfnamefont{S.}~\bibnamefont{Sarkar}},
  \bibinfo{journal}{JHEP} \textbf{\bibinfo{volume}{08}} (\bibinfo{year}{2011}),
  \eprint{1106.3723}, \urlprefix\url{http://arxiv.org/abs/1106.3723}.

\bibitem[{\citenamefont{Lindner et~al.}(2015)\citenamefont{Lindner, Rodejohann,
  and Xu}}]{Lindner:2015iaa}
\bibinfo{author}{\bibfnamefont{M.}~\bibnamefont{Lindner}},
  \bibinfo{author}{\bibfnamefont{W.}~\bibnamefont{Rodejohann}},
  \bibnamefont{and} \bibinfo{author}{\bibfnamefont{X.-J.} \bibnamefont{Xu}}
  (\bibinfo{year}{2015}), \eprint{1510.00666}.

\bibitem[{\citenamefont{Adamson et~al.}(2011{\natexlab{b}})}]{Adamson:2011ku}
\bibinfo{author}{\bibfnamefont{P.}~\bibnamefont{Adamson}} \bibnamefont{et~al.}
  (\bibinfo{collaboration}{MINOS}), \bibinfo{journal}{Phys. Rev. Lett.}
  \textbf{\bibinfo{volume}{107}}, \bibinfo{pages}{011802}
  (\bibinfo{year}{2011}{\natexlab{b}}), \eprint{1104.3922}.

\bibitem[{\citenamefont{Parke and Ross-Lonergan}(2016)}]{sparke:pricom}
\bibinfo{author}{\bibfnamefont{S.}~\bibnamefont{Parke}} \bibnamefont{and}
  \bibinfo{author}{\bibfnamefont{M.}~\bibnamefont{Ross-Lonergan}}
  (\bibinfo{year}{2016}), \bibinfo{note}{private communication}.

\bibitem[{\citenamefont{Conrad et~al.}(2013)\citenamefont{Conrad, Ignarra,
  Karagiorgi, Shaevitz, and Spitz}}]{Conrad:2012qt}
\bibinfo{author}{\bibfnamefont{J.~M.} \bibnamefont{Conrad}},
  \bibinfo{author}{\bibfnamefont{C.~M.} \bibnamefont{Ignarra}},
  \bibinfo{author}{\bibfnamefont{G.}~\bibnamefont{Karagiorgi}},
  \bibinfo{author}{\bibfnamefont{M.~H.} \bibnamefont{Shaevitz}},
  \bibnamefont{and} \bibinfo{author}{\bibfnamefont{J.}~\bibnamefont{Spitz}},
  \bibinfo{journal}{Adv. High Energy Phys.} \textbf{\bibinfo{volume}{2013}},
  \bibinfo{pages}{163897} (\bibinfo{year}{2013}), \eprint{1207.4765}.

\bibitem[{\citenamefont{Honda et~al.}(2004)\citenamefont{Honda, Kajita,
  Kasahara, and Midorikawa}}]{ref:Honda}
\bibinfo{author}{\bibfnamefont{M.}~\bibnamefont{Honda}},
  \bibinfo{author}{\bibfnamefont{T.}~\bibnamefont{Kajita}},
  \bibinfo{author}{\bibfnamefont{K.}~\bibnamefont{Kasahara}}, \bibnamefont{and}
  \bibinfo{author}{\bibfnamefont{S.}~\bibnamefont{Midorikawa}},
  \bibinfo{journal}{Phys. Rev. D} \textbf{\bibinfo{volume}{70}},
  \bibinfo{pages}{043008} (\bibinfo{year}{2004}).

\bibitem[{Ice()}]{IceDataRelease}
\bibinfo{howpublished}{\url{http://icecube.wisc.edu/science/data/IC86-sterile-neutrino}}.

\bibitem[{\citenamefont{Jones}(2015)}]{Jones:2015bya}
\bibinfo{author}{\bibfnamefont{B.~J.~P.} \bibnamefont{Jones}}, Ph.D. thesis,
  \bibinfo{school}{MIT} (\bibinfo{year}{2015}),
  \urlprefix\url{http://lss.fnal.gov/archive/thesis/2000/fermilab-thesis-2015-17.pdf}.

\bibitem[{\citenamefont{Arguelles~Delgado}(2015)}]{Arguelles:2015}
\bibinfo{author}{\bibfnamefont{C.~A.} \bibnamefont{Arguelles~Delgado}}, Ph.D.
  thesis, \bibinfo{school}{University of Wisconsin - Madison}
  (\bibinfo{year}{2015}),
  \urlprefix\url{http://search.proquest.com/docview/1720322773}.

\end{thebibliography}






\newpage

\ifx \standalonesupplemental\undefined
\setcounter{page}{1}
\setcounter{figure}{0}
\setcounter{table}{0}
\fi

\newcolumntype{L}[1]{>{\arraybackslash}p{#1}}
\newcolumntype{C}[1]{>{\centering\arraybackslash}p{#1}}
\newcolumntype{R}[1]{>{\hfill\arraybackslash}p{#1}}

\renewcommand{\thepage}{Supplementary Methods and Tables -- S\arabic{page}}
\renewcommand{\figurename}{SUPPL. FIG.}
\renewcommand{\tablename}{SUPPL. TABLE}


\section*{Appendix:  Incorporating the IceCube Data Into the Fits \label{Icefits}}

The fits proceed by first obtaining the prediction for IceCube for a given 3+1 model.  To obtain this, 
we use the Honda-Gaisser \cite{ref:Honda} prediction for the unoscillated flux.
The nuSQuIDS package \cite{nusquids} is used to propagate the neutrino flux across the Earth.  The software solves the neutrino evolution ``master equation" that accounts for absorption, regeneration and oscillations (Eqs. (29-30) in Ref. \cite{GonzalezGarcia:2005xw}).  
A propagated flux hypothesis is used to produce a weight for each neutrino energy, zenith angle and flavor at the detector.  This can then be used to re-weight the IceCube simulated events available in Ref. \cite{IceDataRelease}. 

The IceCube likelihood is given by \cite{Jones:2015bya, Arguelles:2015, IceCubePRL}:
\small\begin{align}
\ln \mathcal{L}(\vec\theta) = \mathrm{min}_{\eta} \bigg(& \sum^{N_{bins}}_{i=1} \left[ x_i \ln \mu_i(\vec{\theta};\eta) - \mu_i(\vec{\theta},\eta) - \ln x_i! \right] \nonumber \\
& + \frac{1}{2} \sum_\eta \frac{(\eta - \bar\eta)^2}{\sigma^2_\eta}\bigg),  \label{LLIce}
\end{align}\normalsize
where $x_i$ are the number of events in the ith bin, $\mu_i$ is the MC expectation in the same bin, given nuisance parameters ($\eta$) and oscillation parameters ($\vec\theta$). 

We include the nuisance parameters specified in \cite{IceDataRelease,Jones:2015bya, Arguelles:2015, IceCubePRL}. 
We maximize the likelihood as a function of flux variants at each parameter point.  We note this is an important step in reproducing the IceCube limit.

In Supplementary Figure~\ref{fig:appendixth24}, we show the IceCube 90\% CL limit obtained by this analysis.   As expected, the result for $\theta_{34}=0^\circ$, shown by the solid line, is in agreement with that presented in Refs. \cite{Jones:2015bya, Arguelles:2015, IceCubePRL} for the ``rate+shape'' analysis.   This limit is modified at high $\Delta m^2$ if $\theta_{34}$ is nonzero, as discussed in Ref.~\cite{Lindner:2015iaa}.  We illustrate the effect with the 90\% CL limit for IceCube assuming $\theta_{34}=15^\circ$, shown by the dashed line.   Also overlaid on this plot is the global fit result, including the IceCube result, expressed as a function of $\Delta m^2_{4 1}$ vs $\sin^2{2\theta_{24}}$.

In order to combine the IceCube result with the other experiments in Table 1 of Ref.~\cite{Collin:2016rao}, the IceCube likelihood must be converted to a $\chi^2$ that can be combined with that of the global fit. In order to do this, it is convenient to use:
\begin{equation}
\ln \mathcal{LR}(\vec\theta) = \ln \left( \frac{\mathcal{L}(\sin^2 2 \theta_{24}, \Delta m^2_{41})}{\mathcal{SP}(\{x_i\})} \right),
\label{LLR}
\end{equation}
where $\mathcal{SP}(\{x_i\}) = \prod_i \mathbb{P}_{poisson}\left(x_i |x_i\right)$ is the saturated Poisson. 
In the fitting code, we implement 
 Eq.~\ref{LLIce}, and in  Tab.~\ref{tab:bfpointsIce} we use Eq.~\ref{LLR}, following the same procedure as for LSND and Karmen in Ref.~\cite{Collin:2016rao}. (The definitions in Eq.~\ref{LLIce} and Eq.~\ref{LLR} give the same $\Delta\ln\mathcal{L}$ used for fitting and parameter estimation.)

The high computational cost of propagating neutrino fluxes through the Earth with nuSQuIDS prevents the analysis from being directly included into the global fitting software. Instead, the global fits were used to find a reduced set of parameters (called ``parameter-set points" below) that could be evaluated directly.
The 60,000 parameter-set points for the SBL global fits from Ref.~\cite{Collin:2016rao}, {\it i.e.} without IceCube data included,  with the lowest $\chi^2$ were used.  From these every 40th point was selected. This gave a fine sampling of the global fit near the minima. Of the remaining $\approx$140,000 parameter-set points that are far from the minima, every 60th was selected.  Combining the fine sample and the coarse sample yields 4,000 points. 
These 4,000 selected parameter-set points only explored changes in the values for $\theta_{14}$ and $\theta_{24}$. In order to incorporate the IceCube data and effects of $\theta_{34}$, ten values of the $\theta_{34}$ angle were chosen for each parameter point, resulting in a total of 40,000 parameter-set points. These points were fed into the IceCube analysis likelihood and the resulting $\chi^2$ value, defined by Eq. \ref{LLR}, was combined with the respective frequentist global fit $\chi^2$.   

This assumes that the effect of IceCube on the global fit is a small perturbation.  This is reasonable given that the IceCube-only $\Delta \chi^2$ is small compared to the SBL only global fit $\Delta \chi^2$, as shown in Table \ref{tab:bfpointsIce}.

\begin{figure}[t!]
\center
\includegraphics[width=\columnwidth]{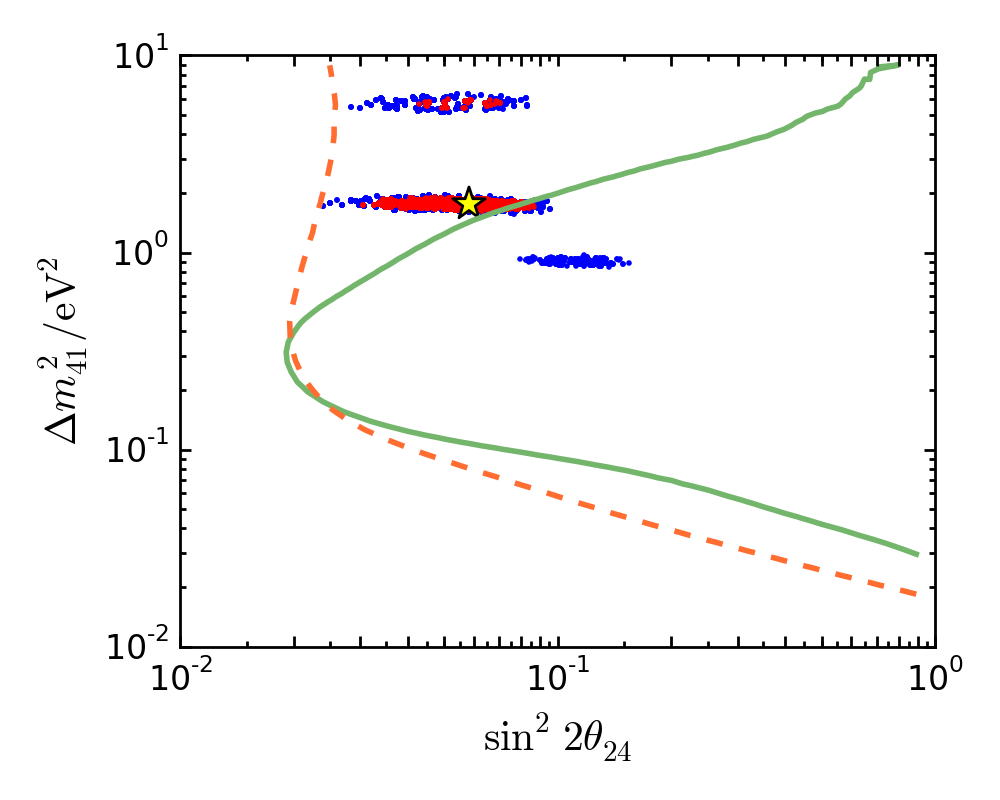}
\caption{The solid (dashed) line represents the 90\% C.L. IceCube limit when calculated with $\theta_{34} = 0\degree$  ($\theta_{34} = 15\degree$).  The result of the SBL+IC global fit is overlaid, Red -- 90\% CL; blue--99\% CL. \label{fig:appendixth24}}
\end{figure}

\clearpage

\end{document}